# Emotion-Guided Image to Music Generation

SOURAJA KUNDU*

Department of Electronics and Electrical Engineering, Indian Institute of Technology Guwahati, India

k.souraja@iitg.ac.in

SAKET SINGH

Visual Solutions Team, Samsung R&D Institute Noida, India

saket.singh@samsung.com

YUJI IWAHORI

Department of Computer Science, Chubu University, Japan

iwahori@isc.chubu.ac.jp

Generating music from images can enhance various applications, including background music for photo slideshows, social media experiences, and video creation. This paper presents an emotion-guided image-to-music generation framework that leverages the Valence-Arousal (VA) emotional space to produce music that aligns with the emotional tone of a given image. Unlike previous models that rely on contrastive learning for emotional consistency, the proposed approach directly integrates a VA loss function to enable accurate emotional alignment. The model employs a CNN-Transformer architecture, featuring pre-trained CNN image feature extractors and three Transformer encoders to capture complex, high-level emotional features from MIDI music. Three Transformer decoders refine these features to generate musically and emotionally consistent MIDI sequences. Experimental results on a newly curated emotionally paired image-MIDI dataset demonstrate the proposed model's superior performance across metrics such as Polyphony Rate, Pitch Entropy, Groove Consistency, and loss convergence.

**ACM Reference Format:**

First Author's Name, Initials, and Last Name, Second Author's Name, Initials, and Last Name, and Third Author's Name, Initials, and Last Name. 2018. The Title of the Paper: ACM Conference Proceedings Manuscript Submission Template: This is the subtitle of the paper, this document both explains and embodies the submission format for authors using Word. In Woodstock '18: ACM Symposium on Neural Gaze Detection, June 03–05, 2018, Woodstock, NY. ACM, New York, NY, USA, 10 pages. NOTE: This block will be automatically generated when manuscripts are processed after acceptance.

---

*Corresponding author

## 1 INTRODUCTION

With the success of Deep Learning and Computer Vision in recent years, the intersection of digital image processing and music generation has become a promising field of research. Both images and music are forms of art that enhance human perception. Traditional digital image processing techniques, such as contrast enhancement [37], deblurring [38], noise removal [39], morphological processing [40], and multiresolution processing [41], improve the visual quality of images. Another effective approach to enhance image perception is transforming images into music. By translating visual features into musical notes, chords, and dynamics, such as a piano piece, a lasting audio-visual impression can be created [42]. This method leverages the strengths of both visual and auditory modalities, enriching the observer's overall experience. This work demonstrates the relevance of digital image processing by extending its application to the auditory domain.

For meaningful music generation, the created music should have contextual similarity with the image, meaning it should provide complementary information. The best way to achieve this relevance is through inducing emotional similarity, ensuring the music evokes similar emotions in the observer as the image does. Verma et al. proposed a deep learning model to predict the affective correspondence between image and music by projecting data from both modalities to a common representation space [43]. Su et al. address this cross-media retrieval problem of aligning photos with music by fuzzily emotionalizing image and music data and then calculating their emotional similarities based on the membership values and emotion confidences [44].

Emotionally relevant image-to-music generation has numerous applications. For example, it can be used in social media to generate background music for photo slides, making them more relatable and memorable. Musicians could create a full piece of music inspired by an image, or instrumental music students could be provided with examples of emotionally relevant songs to learn. Additionally, background music can be added to an image sequence to create a video [1], specific emotional ringtones can be generated using equivalent images, and Guided Imagery and Music (GIM) therapy [2] can employ music to induce a state of relaxed awareness, allowing individuals to revisit pleasant visual imagery for therapeutic benefits [29].

Generative AI can be used for image [45], music [46], text [47], and font synthesis [48]. However, generating emotionally relevant music from images is a complex problem due to the lack of a definite one-to-one mapping between images and emotions. The spectrum of possible emotions an image can evoke cannot be easily categorized into a few discrete emotional classes. Each image has its unique emotional features; for example, two images that both convey happiness can be significantly different. This indicates that emotions are continuous rather than only discrete class labels of 'happy', 'sad', 'calm', or 'angry'. Additionally, different individuals may experience different emotions when viewing the same image, based on their personal experiences and intuition. Therefore, generating appropriate, emotionally relevant music from images is subjective. Deng et al. [36] extracted features such as intensity, timbre, rhythm, pitch, tonality, and harmony to represent music emotions. In cognitive psychology, emotion is well-defined using three parameters: Valence, Arousal, and Dominance (VAD). These parameters provide a continuous representation of emotions. Valence denotes the degree of pleasantness, ranging from positive to negative; Arousal indicates the intensity of emotion, from excited to calm; and Dominance represents the level of control, from submissive to dominant. Due to the difficulty in predicting dominance, most previous works [3,4,5], including this one, use only Valence and Arousal (VA) values to represent emotions. These VA values guide the training process to ensure the generated music is emotionally relevant to the image and help evaluate the emotional similarity between the image and the music.

This work presents a Transformer-based model for generating emotionally relevant music from images, introducing a novel loss function called the Valence-Arousal (VA) loss function to enforce emotional similarity. The MIDI format is used for music representation because it provides better control over musical elements than audio formats. While converting MIDI to other formats is straightforward, converting audio formats like MP3 or WAV to MIDI is not feasible. MIDI sequences are similar to text data, making the image-to-sequence generation model suitable for this task. The contributions are summarized as follows.

- A Transformer-based image-to-sequence model is proposed for generating music with an emotional context similar to the input image. The use of Transformers allows for the creation of rich, contextually relevant music and the generation of long MIDI sequences, which is beneficial for applications requiring extended music synthesis, such as ringtones.
- A novel VA loss function is introduced that enforces emotional similarity between the generated and reference MIDI during training.

The paper is structured as follows: Section 2 discusses related works, Section 3 details the model architecture and algorithm, Section 4 covers the experiments and results, and Section 5 provides the conclusion.

## 2 RELATED WORKS

With the success of deep learning in vision and language tasks, the field of audio and music generation has advanced significantly. Models like MusicLM [6] from Google leverage large-scale language models to generate high-fidelity music from text descriptions through hierarchical sequence-to-sequence modeling. Meta's MusicGen [7] is a single-stage transformer language model that produces high-quality mono and stereo music samples conditioned on textual descriptions or melodic features, using efficient token interleaving patterns to eliminate the need for cascading multiple models. OpenAI's Jukebox [8] introduces a novel hierarchical VQ-VAE architecture combined with autoregressive transformers to generate high-fidelity, long-form music samples with coherent lyrics and musical structure. These industrial initiatives indicate a growing interest in AI-based music generation.

In the literature on audio generation, there are two main approaches: GAN [12]-based models and LLM-based models. Notable GAN-based models include MuseGAN [9], MIDINet [10], and WaveGAN [11]. However, the mode-collapse problem during GAN training limits their widespread acceptance for various tasks. An alternative is autoencoder [13] and transformer [14]-based models like Music Transformer [35], specifically large language models (LLM) like ChatMusician [15]. Music generation in the form of MIDI can be considered text generation, which is why LLM models are well-suited for this task. In multi-modal generation tasks, GPT-4 [16] Omni represents a significant step forward, enabling the generation of coherent audio-visual content. Inspired by the success of transformer-based models, transformers are adapted in the proposed model due to their superior ability to generate contextually relevant music. Existing models often prioritize audio fidelity but may lack the nuanced emotional alignment the proposed approach seeks to address.

The image-to-music generation task presents several challenges. Unlike text-music models trained on substantially large datasets, MIDI data is limited, especially for emotionally paired image-music datasets. Establishing a one-to-one relationship between image and music is difficult for two reasons: music space exhibits a simpler cluster structure than image space, and the relationship between the two spaces is complex and nonlinear [17]. This work aims to overcome these challenges by leveraging transformers to generate emotionally aligned music from images.

Among previous works on image-to-music generation, traditional pipelines translate images to text through image captioning, generate lyrics from captions, and then convert text to music [18]. This approach is computationally heavy because three models should be trained separately. Direct image-to-music generation models on the other hand map visual features, such as color [19], and edges, directly to musical notes. For example, the MuSyFI [20] model generates music by combining shape and position-related features, manual harmony addition, and genetic algorithms. However, these methods rely on low-level visual features and do not account for high-level emotional features like valence arousal (VA) values, which is the focus of this work.

The IM2WAV [21] model generates semantically relevant audio from images by using transformer models conditioned on CLIP (Contrastive Language Image Pre-training) [22] embeddings, with a focus on contextual alignment. In contrast, the proposed approach prioritizes emotional alignment by utilizing a VA loss function, ensuring that the generated music accurately reflects the emotional tone of the image, all within a simpler architecture. The contrastive learning-based continuous emotion-to-music generation model by Wang et al. [29], while versatile, can become complex due to the need for separately training each component. The proposed end-to-end model, however, can be trained in a single process. Drawing inspiration from Mendis et al.'s [23] CNN [24]-LSTM [25] model for melody generation, a CNN-Transformer image-to-sequence model is adapted, which enhances emotional consistency in music generation by integrating pre-trained image encoders and transformer-based MIDI sequence processors.

## 3 METHODOLOGY

This section first discusses the strategy for preparing emotionally paired image-music datasets to train the proposed model. Next, it details the algorithm for image-to-music generation. Finally, it delves into the model architecture and implementation details.

### 3.1 Dataset Preparation

The image-MIDI emotionally paired dataset was created by combining two independent sources of music and images. 4029 images were downloaded from the Emotionally Paired Music and Image Dataset (EMID) [27], tagged with emotions like "anger," "amusement," "awe," and "fear." An emotion-to-VA value mapping dictionary [28] was used to find the corresponding VA values for these images. However, the corresponding audio files were in .mp3 format, which was unsuitable for this work, as MIDI files were required, and converting .mp3 to MIDI is challenging. Instead, a 3000 VA-labeled MIDI dataset from [26] was obtained. To align the VA value ranges of both datasets, they were normalized to the range (1,9). The 3000 MIDI files and 4029 images were then paired based on emotional similarity, determined using the following Similarity Score:

$$s_{xy} = \left(\left(v_x - v_y\right)^2 + \left(a_x - a_y\right)^2\right)^{-1/2}$$

Here, x represents the image, and y represents the music. The variables $v_x$ and $v_y$ are their respective Valence values, while $a_x$ and $a_y$ denote their Arousal values. This process resulted in a dataset of 3000 pairs, which was further divided randomly into 2884 training pairs, 100 testing pairs, and 16 validation pairs.

### 3.2 Model Workflow

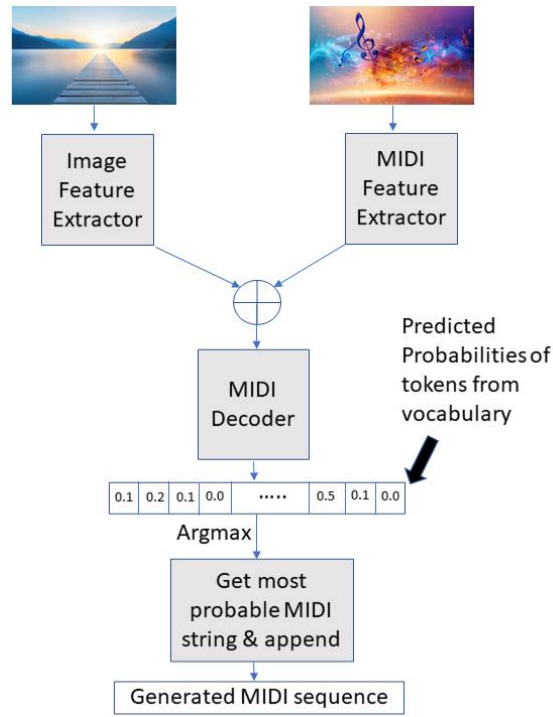

Figure 1: Algorithm for Image to Music Generation

Since MIDI music data is similar to sequential text data, an image-to-sequence model is leveraged for this task. Inspired by Mendis et al. [23], an architecture comprising three key modules is implemented: an image feature extractor, a MIDI feature encoder, and a MIDI decoder. The image feature extractor processes visual input to identify and encode essential features, capturing the emotional and contextual essence of the image. Concurrently, the MIDI feature encoder processes the MIDI data sequence, preserving its sequential nature and musical structure. By merging the outputs of these two encoders, a rich, multimodal feature representation is created that captures both visual and auditory elements. The music decoder then uses this combined representation to generate MIDI sequences that are emotionally and contextually aligned with the original image. A novel loss function called the Valence-Arousal (VA) loss, is implemented to ensure emotional alignment. This loss function enforces emotional similarity between the generated music and the reference music, which is already emotionally paired with the image through similarity score, ensuring that the final output resonates with the intended emotional context. Figure 1 illustrates the algorithm.

### 3.3 Model Architecture

*3.3.1 Image Encoder*

First, the image feature encoder is implemented using pre-trained VGG16 and InceptionV3 models. Given the architecture's plug-and-play nature, other image feature encoders can also be experimented with. For instance, Wang et al. [29] explored Adversarial Latent Auto Encoders (ALAE) [30], Vector Quantized Variational Auto Encoder (VQVAE) [31], and β-Variational Auto Encoder (β-VAE) [32]. These models learn a disentangled representation of image data using various techniques: ALAE combines GANs and autoencoders, VQVAE utilizes vector quantization, and β-VAE encourages disentangled and interpretable data representations through regularization. However, these models require large training datasets for optimal performance.

In contrast, the VGG16 [33] and InceptionV3 [34] models, pre-trained on a subset of ImageNet with 14 million images from 22,000 categories, provide efficient and reliable image encoding. For this implementation, VGG16 is modified by omitting the last two fully connected layers and applying a pooling layer with a kernel size of 3 to the resulting 7x7x512 feature tensor, producing a 1x1x512 image feature vector. This approach ensures effective image feature extraction without extensive additional training.

*3.3.2 MIDI Encoder*

For processing the MIDI text sequence, a word embedding layer is employed, followed by a series of transformer encoder blocks (using 2, 3, or 4 blocks). Each transformer encoder block contains stacked multi-headed attention mechanisms and layer normalization, and finally, there is a global average pooling layer. Unlike LSTMs, transformers can better capture contextual relationships and efficiently transfer the abstract emotional representation of the music.

*3.3.3 MIDI Decoder*

After merging the extracted features from both the image and MIDI encoders, the combined feature is processed through this module. It consists of a varying number of transformer decoder blocks (2 or 3). Each decoder block includes multi-head attention, a flattening layer, layer normalization, and a dense layer with ReLU activation.

*3.3.4 Objective Function*

The last fully connected layer of the MIDI decoder predicts the probabilities of the current MIDI token belonging to the 977 vocabulary tokens. A categorical cross-entropy loss function is employed since this task resembles multi-class classification.

$$\text{LCC} = - \sum_{i=1}^{N} \sum_{j=1}^{C} y_{ij} \log(\hat{y}_{ij})$$

Here, N = maximum MIDI sequence length, C = number of tokens in the vocabulary, $y_{ij}$ is a one-hot encoded vector representing the actual class (note) at the $i^{th}$ time step and $\hat{y}_{ij}$ is the predicted probability of the $j^{th}$ MIDI token (note) at the $i^{th}$ time step, given by the model. However, the generated MIDI must also have the same emotional parameters, or VA values, as the reference image-MIDI pair. Therefore, a VA loss function is introduced in addition to the categorical cross-entropy loss.

The VA loss function works as follows: first, the true and predicted MIDI token sequences are obtained by performing an Argmax operation on each column of the probability tensors (of size vocabulary length × maximum sequence length). These token IDs are then converted to MIDI strings. A pre-trained music VA predictor, consisting of 3 fully connected layers (each followed by BatchNorm and ReLU activation layers, except for the last output layer which uses a linear activation function) [29], predicts the VA values of both the true and generated MIDI sequences. The Mean Absolute Error (MAE) between these VA values is calculated as the VA loss.

$$L_{VA} = \frac{1}{N}\sum_{i=1}^{N} |VA_{true}^{i} - VA_{pred}^{i}|$$

The final objective function of the model is:

$$L_{total} = \lambda_1 \cdot L_{VA} + \lambda_2 \cdot L_{CC}$$

where $\lambda_1 = 0.00001$ and $\lambda_2 = 1$ are chosen experimentally to ensure that the overall loss decreases with each epoch.

Figure 2 illustrates the model architecture.

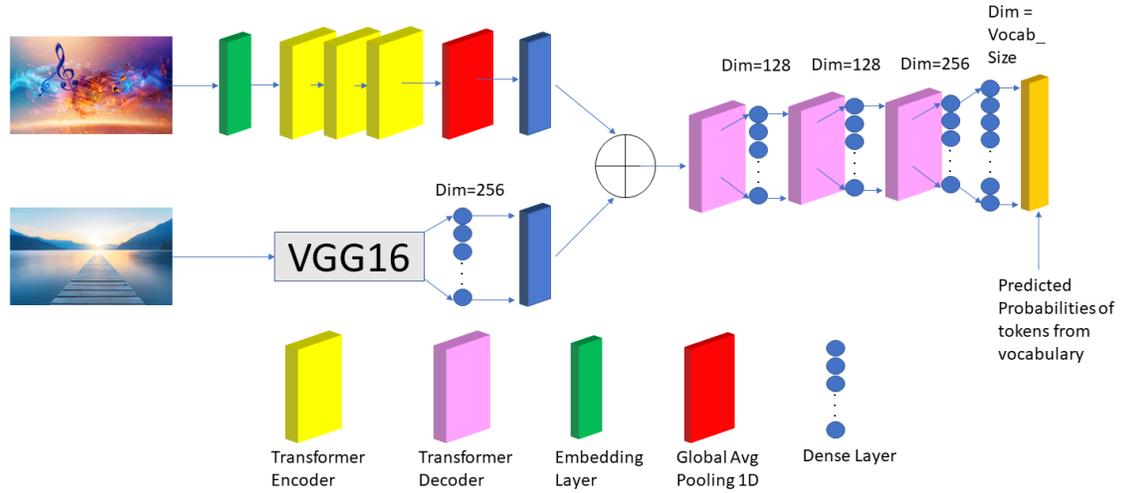

Figure 2: Proposed model architecture

### 3.4 Implementation Details

For a fair comparison, the proposed model along with all its ablated versions was trained via the Adam optimizer with β1 = 0.9, β2 = 0.999. Different learning rates were considered in the set {$10^{-6}$, $10^{-5}$, $10^{-4}$, $2\times10^{-4}$, $2\times10^{-3}$} and the number of epochs in the set {2, 5, 10, 100, 150, 250}. The final learning rate was fixed at $10^{-5}$ and the number of epochs at 15. Modeling was performed via the TensorFlow framework on a single Nvidia RTX A6000 GPU. In the objective function, the $\lambda_1$ values were chosen from the set {1, 0.1, $10^{-2}$, $10^{-3}$, $10^{-4}$, $10^{-5}$}, and $\lambda_2$ values were chosen from the set {1, 0.9, 0.99}. Since the VA loss values for the initial few epochs of training were high, so finally the $\lambda_1$ value was fixed at 0.00001.

### 3.5 Baseline Model

To evaluate the effectiveness of the proposed model with the state-of-the-art similar image-to-music synthesis models, the model proposed in [23] was chosen as the baseline model. In this paper by Mendis et al., they use a model similar to the proposed model but with LSTM and without considering emotional similarity. The Image Feature Extractor utilizes the VGG16 model (excluding the final output layer) to pre-process images and extract feature representations of the input images. The Sequence Processor incorporates a word embedding layer to process textual input derived from MIDI data, followed by two Long Short-Term Memory (LSTM) layers. The Decoder merges the fixed-length vectors outputted by both the image feature extractor and the sequence processor. This combined representation is then processed by a Dense layer, which generates the final musical prediction.

## 4 EXPERIMENTS AND RESULTS

In this section, the experimental results for model comparison and ablation experiments will be shown and their implications will be discussed.

### 4.1 Evaluation Metrics

First, the quality of the generated was evaluated using three metrics as in the paper [23], namely Pitch Entropy, Polyphony Rate, and Groove Consistency.

#### 4.1.1 Pitch Entropy

Pitch Entropy is a measure of the variation or unpredictability of pitches in a piece of music. It tells how diverse or complex the melody is in terms of the range of different notes used. A higher Pitch Entropy indicates that the music has a wide variety of pitches, making it less predictable, while a lower value suggests that the music uses fewer pitches, making it more repetitive.

$$Pitch\ Entropy = -\sum_{i=0}^{127} P(pitch = i)\ log_2\ P(pitch = i)$$

$P(pitch = i)$ is the probability of the i[th] pitch occurring in the music, where $i$ can belong to the range 0 to 127 of all possible MIDI pitches.

#### 4.1.2 Polyphony Rate

Polyphony Rate measures how often multiple notes are played simultaneously in a piece of music. It's an indicator of the musical texture. A higher Polyphony Rate means the music is more complex, with several notes being played together (e.g., chords). In comparison, a lower rate suggests simpler, monophonic music (e.g., a single melody line).

$$Polyphony\ Rate = \frac{MPA\_time\_steps}{time\_steps}$$

MPA_time_steps is the number of time steps out of total time_steps during which multiple pitches are played simultaneously.

*4.1.3 Groove Consistency*

Groove Consistency evaluates how stable and consistent the rhythm is throughout the music. It focuses on the regularity of rhythmic patterns and timing, aiming to measure how well the rhythm sticks to a consistent beat. A higher Groove Consistency indicates a steady rhythm, while a lower value suggests more variability or inconsistency in the rhythm.

$$Groove\ Consistency = 1 - \frac{1}{T-1}\sum_{i=1}^{T-1} d(g_i, g_{i+1})$$

T is the total number of measures (sections of a musical piece). $g_i$ is the binary onset vector for the i[th] measure, where a "1" indicates a note onset (when a note starts) and a "0" indicates no onset. $d(g_i, g_{i+1})$ is the Hamming distance between the onset vectors of consecutive measures, which measures how similar or different the rhythms of these two measures are.

**4.2 Ablation Study**

For the ablation study, the model was first tested with two pre-trained image feature extractors, VGG_16 and Inception_V3, and based on the metrics, VGG_16 was finalized. Then different numbers of transformers were experimented with in the encoder and the decoder, including models with only dense layers in the decoder. Results showed that fewer than 3 transformers failed to capture complex emotional features, generating redundant tones in the same music and different images mapping to the same music, while more than 3 transformers again led to repetitiveness in generated tones due to the limited dataset size compared to the number of parameters. The optimal performance was achieved with 3 transformers in both the encoder and decoder, balancing complexity and data efficiency. The ablation study also clarified the role of transformer decoders: they enrich representations with contextual information, allowing dense layers to work with more refined, context-aware features, thus enhancing model performance. Additionally, increasing the number of transformers in the encoder will prove beneficial only with a larger dataset size. In Table 1, the Music_Quality_Loss value is the average loss with respect to the ground truth values of the 3 following music quality metrics, i.e. [0.5303, 3.9863, 0.9922] after 2 epochs, since the trend in loss convergence was apparent even with training for a smaller number of epochs. In Table 2, the comparison of music quality of music generated by models with and without the VA loss is shown. It should be noted that objective evaluation using these music quality metrics is not always reliable for all different generated music, in those cases subjective evaluation was prioritized. Also, the same image can be mapped to different music from the same model, it is because there is a one-to-many relationship between the image and music.

| Table 1: Ablation Experiments (after 2 epochs) | | | | |
|---|---|---|---|---|
| Model | Music_Quality_Loss | Polyphony Rate | Pitch Entropy | Groove Consistency |
| LSTM_enc2 (baseline) | 1.9157 | 0.0317 | 0.1204 | 0.9978 |
| Trans_enc3_dec3_VGG | **0.8216** | **0.6975** | **1.9974** | 0.9995 |
| Trans_enc3_dec3_Inception | 0.8514 | 0.6289 | 1.3642 | 0.9998 |
| Trans_enc4 | 1.3368 | 0.0905 | 1.9931 | 0.9997 |
| Trans_enc2_dec2 | 2.0077 | 0.0000 | 0.0000 | **0.9999** |

| Table 2: Validity of VA loss (after 15 epochs) | | | | |
|---|---|---|---|---|
| Model | Music_Quality_Loss | Polyphony Rate | Pitch Entropy | Groove Consistency |
| Trans_enc3_dec3 | 1.2853 | 0.0087 | 2.8137 | 0.9997 |
| Trans_enc3_dec3_VA_Loss | **0.7061** | **0.7818** | **3.0932** | **0.9998** |

### 4.3 Model Comparison

The proposed model was compared to the LSTM-based baseline model and metrics as given in Table 1 validate the proposed model. Figure 3 shows the total loss value vs epoch plot for the baseline model LSTM_enc2 (2 LSTMs in the encoder) and the proposed model trans_enc3_dec3_VGG_VA_Loss (3 transformers in both encoder and decoder with the VA loss function) for 15 epochs. It can be seen that for the proposed model, the loss function converges much faster. For the LSTM model, the total loss function includes only categorical cross-entropy loss, while for the proposed model, the net loss function includes both the categorical cross-entropy and the VA loss function.

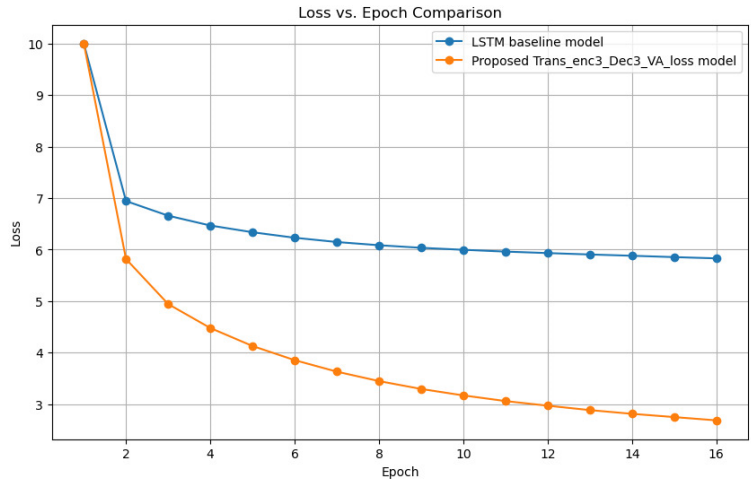

Figure 3: Loss vs Epoch plot comparison of proposed vs baseline model

### 5 FUTURE WORK AND CONCLUSION

In this paper, it is shown that implementing transformers instead of LSTMs in an image-to-sequence task-inspired model can significantly improve model performance because Transformers, unlike LSTMs, process the music as a whole with their multi-headed attention rather than sequentially and capture the overall emotional context. Also, the inclusion of the transformer decoder is useful; just like in GAN models, the discriminator should be more powerful than the generator, and in the vanilla transformer architecture, the decoder is heavier than the encoder; similarly, a symmetrical number of transformers in the decoder as in the encoder along with a few more dense layers can capture the intricate high-level emotional features in the generated music more effectively. Several observations have been made regarding the performance of the proposed model: when working with limited paired datasets, reducing the number of transformers in the encoder and decoder can lead to faster convergence, though this may increase the risk of overfitting. For instance, using trans_enc2 (with only 2 transformers in the encoder) has shown significant improvements in as few as 2 epochs, but the learning was not generalized. If emotional alignment is not a priority, the VA loss can be omitted, which further accelerates training because the current implementation of VA loss saves and reads the MIDI files for each image-MIDI pair during the training, taking slightly longer training time, the future goal is to optimize the algorithm of this loss function evaluation. Additionally, if a smaller model size is desired, the transformer in the decoder can be removed, with the number of transformers in the encoder adjusted through validation to optimize test performance. However, this might degrade the model's capability to generate rich varieties of music.

Future work will focus on expanding the dataset to include multi-instrumental and vocal music, beyond just single piano compositions, to broaden the applicability of the proposed model. It is hoped that these insights contribute to further advancements in the domain of image-to-music generation with emotional alignment.

**ACKNOWLEDGMENTS**

This research was conducted when Souraja Kundu visited the Samsung R&D Institute in Noida, India from May to July 2024. This paper is partially supported by the Chubu University grant.

# Authors' background

| Your Name | Title* | Research Field | Personal website |
|---|---|---|---|
| **Souraja Kundu** | **Bachelor's student** | **Computer Vision, NLP, Audio** | |
| **Saket Singh** | **R&D Engineer (Samsung R&D Institute India)** | **Audio** | |
| **Yuji Iwahori** | **Full Professor** | **Computer Vision** | |